\documentclass{epl}
\usepackage{amsmath}

\newcommand{\llangle}{\langle\!\langle}
\newcommand{\rrangle}{\rangle\!\rangle}
\newcommand{\rlangle}{\rrangle\!\llangle}

\title{Full counting statistics of nano-electromechanical systems}
\shorttitle{FCS of NEMS}
\author{Christian Flindt\inst{1}\thanks{E-mail: \email{cf@mic.dtu.dk}} \and Tom\'{a}\v{s} Novotn\'{y}\inst{1,2} \and Antti-Pekka Jauho\inst{1}}
\institute{
  \inst{1} MIC -- Department of Micro and Nanotechnology, Technical University of Denmark,
  DTU - Building 345east, DK-2800 Kongens Lyngby, Denmark\\
  \inst{2} Department of Electronic Structures, Faculty of
  Mathematics and Physics, Charles University - Ke Karlovu 5, 121
  16 Prague, Czech Republic
} 

\begin{document}

\maketitle

\begin{abstract}
We develop a theory for the full counting statistics (FCS) for a
class of nanoelectromechanical systems (NEMS), describable  by a
Markovian generalized master equation.  The theory is applied to
two specific examples of current interest: vibrating \chem{C_{60}}
molecules and quantum shuttles. We report a numerical evaluation
of the first three cumulants for the ${\rm C}_{60}$-setup; for the
quantum shuttle we use the third cumulant to substantiate that the
giant enhancement in noise observed at the shuttling transition is
due to a slow switching between two competing conduction channels.
Especially the last example illustrates the power of the~FCS.
\end{abstract}

\section{Introduction}

The full counting statistics (FCS) of charge transport in
mesoscopic systems is an active topic of recent
research\cite{levitov_1996,nazarov_2003, pilgram_2003,
bagrets_2003,roche_2003}. Calculation and measurement of the whole
probability distribution of transmitted charge is motivated by the
fact that FCS provides more information about a particular system
than just the mean current or current noise which are the first
two cumulants of the large-time asymptotics of the probability
distribution. Very recently, a measurement of the third cumulant,
which quantifies the skewness of the distribution, was
reported\cite{reulet_2003}. The detailed nature of charge
transport in nanoelectromechanical systems (NEMS), another modern
field in mesoscopics, poses many challenges both to experiments
and theory, and the computation of FCS for NEMS is an important
task that needs to be addressed. The first steps were taken
recently with a calculation of FCS for a {\it driven, classical}
shuttle\cite{pistolesi_2004}.

In this Letter, we present a theory for the evaluation of
cumulants in a wide class of NEMS encompassing the majority of
systems considered thus far, namely those which can be described
by a Markovian generalized master equation (GME). The current
cumulants turn out to be fully determined by an extremal
eigenvalue of the system evolution superoperator (Liouvillean) in
analogy with previous studies\cite{bagrets_2003, roche_2003}.
Their evaluation is, however, more complicated since in NEMS there
are generally many relevant states which need to be taken into
account. We solve the problem by formulating a systematic
perturbation theory, and using this derive explicit formulas for
the first three cumulants. The method is illustrated by two
examples of NEMS -- the \chem{C_{60}}-experiment\cite{park_2000}
and the quantum
shuttle\cite{novotny_2003,novotny_2004,fedorets_2004}. To test the
method we calculate the first three cumulants for the model of the
\chem{C_{60}}-setup from\cite{mccarthy_2003}. In case the
oscillator is equilibrated the cumulants  can be calculated
alternatively using $P(E)$-theory which gives the same results.
For the quantum shuttle we use the third cumulant to substantiate
that the giant enhancement of the current noise in the transition
region\cite{novotny_2004} is caused by a slowly fluctuating
amplitude of the shuttle resulting in a slow switching between two
current channels, \emph{i.e.} tunneling and shuttling. This part
complements\cite{pistolesi_2004}, which considered a fixed driving
amplitude, and\cite{isacsson_2004} describing a related phenomenon
in a different model.

\section{Theory}

We consider a nanoelectromechanical system with discrete energy
levels electronically coupled to two leads and mechanically
coupled to a generic heat bath providing dissipation. The system
is described by the reduced density operator $\hat{\rho}(t)$,
which we assume evolves according to the Markovian GME
\begin{equation}
\dot{\hat{\rho}}(t)=\mathcal{L}\hat{\rho}(t).
\end{equation}
The Liouvillean $\mathcal{L}$ describes the dynamics of the system
and we assume that the system tends exponentially to a stationary
state $\hat{\rho}^{\ab{stat}}$. This implies that the Liouvillean,
which is a non-hermitian operator, has a single eigenvalue equal
to zero with $\hat{\rho}^{\ab{stat}}$ being the corresponding
(normalized and unique) right eigenvector which we denote by
$|0\rrangle$ \cite{flindt_2004}. The corresponding left
eigenvector is the identity operator $\hat{1}$, denoted by
$\llangle\tilde{0}|$, and we have
$\llangle\tilde{0}|0\rrangle\equiv\tx{Tr}(\hat{1}^{\dagger}\hat{\rho}^{\ab{
stat}})=1$. The pair of eigenvectors allows us to define the
projectors $\mathcal{P}\equiv|0\rlangle\tilde{0}|$ and
$\mathcal{Q}\equiv1-\mathcal{P}$ obeying the relations
$\mathcal{PL}=\mathcal{LP}=0$ and $\mathcal{QLQ}=\mathcal{L}$. We
also introduce the pseudoinverse of the Liouvillean
$\mathcal{R}\equiv\mathcal{Q}\mathcal{L}^{-1}\mathcal{Q}$, which
is well-defined, since the inversion is performed only in the
subspace spanned by $\mathcal{Q}$, where $\mathcal{L}$ is regular.
The assumption of exponential decay to the stationary state is
equivalent to the spectrum of $\mathcal{L}$ in the subspace
spanned by $\mathcal{Q}$ having a finite negative real part.

In order to evaluate the FCS of the system, \emph{i.e.} the
probability $P_{n}(t)$ of $n$ electrons being collected in, say,
the right lead in the time span $t$, we resolve the density
operator $\hat{\rho}(t)$ and the GME with respect to $n$. The GME
is a continuity equation for the probability (charge) and,
therefore, we can identify terms corresponding to charge transfer
processes between the system and the right lead. Specifically, we
introduce the superoperator $\mathcal{I}^{+}$ of the particle
current of electrons tunneling from the system to the right lead,
and the corresponding superoperator $\mathcal{I}^{-}$ of the
reverse process, where electrons tunnel from the right lead to the
system. In terms of these superoperators the $n$-resolved GME can
be written as
\begin{equation}
\dot{\hat{\rho}}^{(n)}(t)=(\mathcal{L}-\mathcal{I}^{+}-\mathcal{I}^{-})\hat{\rho}^{(n)}(t)
+\mathcal{I}^{+}\hat{\rho}^{(n-1)}(t)+\mathcal{I}^{-}\hat{\rho}^{(n+1)}(t)
\end{equation}
with $n=\ldots,-1,0,1,\ldots$. From the $n$-resolved density
operator we can obtain, at least in principle, the complete
probability distribution
$P_n(t)=\tx{Tr}\left[\hat{\rho}^{(n)}(t)\right]$.

It is practical first to evaluate the cumulant generating function
$S(t,\chi)$ defined as
\begin{equation}
e^{S(t,\chi)}=\sum_{n=-\infty}^{\infty}P_n(t)e^{in\chi}.
\end{equation}
From $S(t,\chi)$ we then find the $m$'th cumulant of the charge
distribution (we take $e=1$)  by taking the $m$'th derivative with
respect to the counting field $\chi$ at $\chi=0$, \emph{i.e.}
$\llangle
n^m\rrangle(t)=\frac{\partial^mS}{\partial(i\chi)^m}|_{\chi=0}$,
and from the knowledge of all cumulants we can reconstruct
$P_n(t)$. The cumulants of the current in the stationary limit
$t\to\infty$ are given by the time derivative of the charge
cumulants, \emph{i.e.} $\llangle I^m\rrangle=\frac{d}{dt}\llangle
n^m\rrangle(t)\big|_{\, t\to\infty}$. The first two current
cumulants give the average current and the zero-frequency current
noise, respectively.

Using $\hat{\rho}^{(n)}(t)$ we may express $S(t,\chi)$ as
$e^{S(t,\chi)}=\tx{Tr}\bigl[\,\sum_{n=-\infty}^{\infty}\hat{\rho}^{(n)}(t)e^{in\chi}\bigr]
=\tx{Tr}\bigl[\hat{F}(t,\chi)\bigr]$, where we have introduced the
auxiliary operator $\hat{F}(t,\chi)$ whose equation of motion
follows from the $n$-resolved GME,
\begin{equation}
\frac{\partial}{\partial t
}\hat{F}(t,\chi)=[\mathcal{L}-(1-e^{i\chi})\mathcal{I}^{+}´-(1-e^{-i\chi})\mathcal{I}^{-}]\hat{F}(t,\chi)\equiv
\mathcal{L}_{\chi}\hat{F}(t,\chi)
\label{eq_eom}
\end{equation}
with the formal solution
$\hat{F}(t,\chi)=e^{\mathcal{L}_{\chi}t}\hat{F}(0,\chi)$. We
assume adiabatic evolution of the spectrum of $\mathcal{L}_{\chi}$
with increasing $\chi$, \emph{i.e.} there is a unique eigenvalue
$\Lambda^{\ab{min}}_{\chi}$ of $\mathcal{L}_{\chi}$ associated
with the projector $\mathcal{P}_{\chi}$ which develops from the
zero eigenvalue of $\mathcal{L}$ and which is the closest to zero
for small enough $\chi$. The rest of the spectrum still has finite
negative real part which ensures the damping of its contribution
for large times. Thus, we have
\begin{equation}
e^{S(t,\chi)}=\llangle\tilde{0}|e^{\mathcal{L}_{\chi}t}|F(0,\chi)\rrangle\rightarrow
e^{\Lambda^{\ab{min}}_{\chi}t}\llangle\tilde{0}|\mathcal{P}_{\chi}|F(0,\chi)\rrangle
=e^{\Lambda^{\ab{min}}_{\chi}t+C_{\chi}^{\rm init}} \mbox{
\tx{for}
$t\rightarrow\infty$,}
\end{equation}
where $C_{\chi}^{\rm init}$ depends on the initial state of the
system. However, the current cumulants in the stationary state do
not depend on the initial conditions, but are totally determined
by $\Lambda^{\ab{min}}_{\chi}$ in full analogy with previous
studies \cite{bagrets_2003,roche_2003}. For NEMS in general
$\mathcal{L}_{\chi}$ is of very large dimensions, and the
numerical evaluation of higher order derivatives of
$\Lambda^{\ab{min}}_{\chi}$ may become a formidable numerical
problem. In order to circumvent this problem we determine
$\Lambda^{\ab{min}}_{\chi}$ using Rayleigh-Schr\"{o}dinger
perturbation theory for
$\mathcal{L}_{\chi}=\mathcal{L}+\mathcal{L}'_{\chi}$, treating
$\mathcal{L}'_{\chi}$ as the perturbation. Since the Liouvillean
is not hermitian, we cannot assume that it has a spectral
decomposition in terms of its eigenvectors, and one cannot use
directly the standard formulas. However, it is possible to
formulate the perturbation theory exclusively in terms of the
projectors $\mathcal{P},\,\mathcal{Q}$ and the pseudoinverse
$\mathcal{R}$. As in standard Rayleigh-Schr\"{o}dinger
perturbation theory the first order correction is given by the
average of the perturbation with respect to the unperturbed
eigenstate. Taking the derivative of the first order correction
with respect to $i\chi$ and letting $\chi\rightarrow 0$ we find
\begin{equation}
\llangle I\rrangle = \llangle\tilde{0}|\mathcal{I}\,|0\rrangle,
\label{eq_firstcumulant}
\end{equation}
where $\mathcal{I}\equiv \mathcal{I}^{+}-\mathcal{I}^{-}$. As
expected the first cumulant equals the average current. For the
second cumulant, \emph{i.e.} the zero-frequency current noise, one
finds
\begin{equation}
\llangle I^2\rrangle =
\llangle\tilde{0}|\mathcal{J}|0\rrangle-2\llangle\tilde{0}|\mathcal{I}\mathcal{R}\mathcal{I}\,|0\rrangle,
\label{eq_secondcumulant}
\end{equation}
where $\mathcal{J}\equiv \mathcal{I}^{+}+\mathcal{I}^{-}$. In the
high bias limit (where
$\llangle\tilde{0}|\mathcal{I}^{-}|0\rrangle=0$, since backward
tunneling is blocked) this expression yields the result previously
derived in\cite{flindt_2004}. The expression for the third
cumulant
\begin{equation}
\llangle I^3\rrangle = \llangle\tilde{0}|\mathcal{I}\,|0\rrangle
-3\llangle\tilde{0}|\mathcal{I}\mathcal{R}\mathcal{J}+\mathcal{J}\mathcal{R}\mathcal{I}\,|0\rrangle
-6\llangle\tilde{0}|\mathcal{I}\mathcal{R}(\mathcal{R}\mathcal{I}\mathcal{P}-\mathcal{I}\mathcal{R})\mathcal{I}\,|0\rrangle
\label{eq_thirdcumulant}
\end{equation}
is the main result of this section, and below we evaluate it for
two specific cases. Higher order cumulants can be obtained in the
same manner by calculating the corresponding higher order
corrections.

\section{Model 1: The \chem{C_{60}} experiment}

An experiment with a NEMS that has received much attention is the
measurement of the $IV$-curves of a vibrating
\chem{C_{60}}-molecule\cite{park_2000}. The experiment has been
modelled in several
papers\cite{boese_2001,mccarthy_2003,mitra_2004} using a model
which will also be employed here. Calculations of
$IV$-curves\cite{boese_2001,mccarthy_2003} have been found to be
in good agreement with the experiment, and the current noise has
been predicted\cite{mitra_2004}. We calculate the third cumulant
for this setup by applying our method to the model as described
in\cite{mccarthy_2003}.

\begin{figure}
  \centering
  \includegraphics[width=0.95\textwidth]{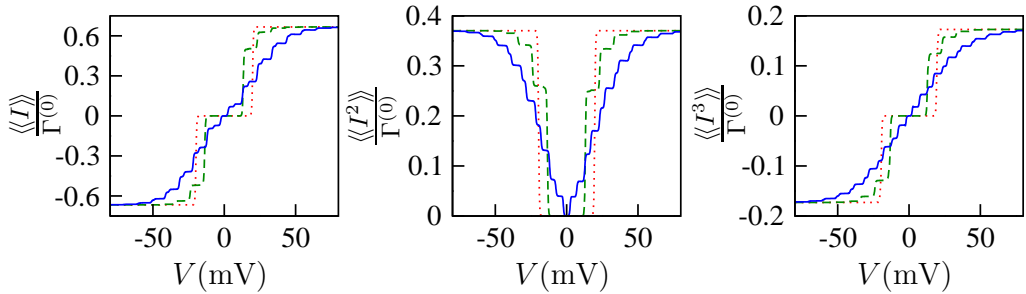}
  \caption{Results for the \chem{C_{60}}-setup. First three cumulants
   as function of the bias $V$. The parameters, which correspond to
   fig.\ 3 of\cite{mccarthy_2003},
   are $E_c = 10 \un{meV},\, \hbar\omega_0 = 5 \un{meV},\, \hbar\Gamma^{(0)}=1 \un{\mu eV},
   \, k_B T= 0.15 \un{meV}$, $K=0.1\omega_0$, and
   $c_1 =0$ (dashed), 0.8 (dotted), 1.5 (full), $c_2=0.005$.}
  \label{fig_resultsC60}
\end{figure}

In the model of\cite{mccarthy_2003} both the coupling to the leads
and to the heat bath are treated in the weak coupling
approximation which reduces the full GME to an ordinary Pauli
master equation for the probabilities of occupation of the
individual eigenstates of the system:
\begin{equation}
\begin{split}
\frac{dP_{m,\sigma,l}}{dt}=&-\Big[W_{l+1\leftarrow l}+W_{l-1\leftarrow
l}+\sum_{m',\sigma',l',s}\Gamma_{m',\sigma',l'\leftarrow
m,\sigma,l}^{(s)}\Big]P_{m,\sigma,l}\\
&+\sum_{m',\sigma',l',s}\Big[\Gamma_{m,\sigma,l\leftarrow
m',\sigma',l'}^{(s)}+W_{l\leftarrow
l'}(\delta_{l+1,l'}+\delta_{l-1,l'})\delta_{\sigma,\sigma'}\delta_{m,m'}\Big]P_{m',\sigma',l'},
\label{eq_modelC60}
\end{split}
\end{equation}
where $m,\sigma,l$ denote the (extra) charge on the molecule
($m=0,1$), the spin, and the vibrational state, respectively,
while $s$ indicates whether an electron tunnelled from/to the left
($s=-1$) or right ($s=1$) lead. $P_{m,\sigma,l}$ is the
probability of being in the eigenstate labelled by the subindices.
Bath-mediated transitions between different vibrational states are
given by the thermal rates
\begin{equation}
W_{l+1\leftarrow l}=W_{l\leftarrow
l+1}e^{-\hbar\omega_0/k_BT}=K\frac{l+1}{e^{\hbar\omega_0/k_BT}-1},
\end{equation}
where $\omega_0$ is the natural oscillator frequency. The charge
transfer rates are
\begin{equation}
\begin{split}
\Gamma_{1,\sigma,l'\leftarrow 0,0,l}^{(s)} &= \Gamma^{(0)}|\langle
l'|e^{\gamma (\hat{a}^{\dagger}-\hat{a})}|l\rangle
|^2f(E_C+\frac{seV}{2}+\hbar\omega_0 (l'-l-\gamma^2)),
\\
\Gamma_{0,0,l\leftarrow 1,\sigma,l'}^{(s)} &= \Gamma^{(0)}|\langle
l|e^{\gamma (\hat{a}^{\dagger}-\hat{a})}|l'\rangle
|^2[1-f(E_C+\frac{seV}{2}+\hbar\omega_0 (l'-l-\gamma^2))],
\end{split}
\end{equation}
where $f$ is the Fermi function, $\Gamma^{(0)}$ is the bare
tunneling rate, $E_C$ is the charging energy difference, and $V$
is the symmetrically applied bias. The quantity $\gamma$ describes
the bias-dependence of the electric field at the position of the
molecule, and is assumed to have the form $\gamma =
c_1+\frac{eV}{\hbar\omega_0}c_2$\cite{mccarthy_2003}. Here we do
not consider the case where the rates depend on the position of
the molecule, although this can easily be included.

The current superoperators are identified from the expression for
the stationary current
\begin{equation}
I^{\ab{stat}}=\underbrace{\sum_{\sigma,l,l'}\left[\Gamma_{0,0,l'\leftarrow
1,\sigma,l}^{(1)}P^{\ab{stat}}_{1,\sigma,l}\right]}_{\llangle\tilde{0}|\mathcal{I}^{+}|0\rrangle
}-\underbrace{\sum_{\sigma,l,l'}\left[\Gamma_{1,\sigma,l'\leftarrow
0,0,l}^{(1)}P^{\ab{stat}}_{0,0,l}\right]}_{\llangle\tilde{0}|\mathcal{I}^{-}|0\rrangle
}.
\end{equation}
Here $|0\rrangle$ is a diagonal density matrix containing the
stationary probabilities $P^{\ab{stat}}_{m,\sigma,l}$. Due to this
diagonal form of the density matrices the relevant superoperators
needed for the cumulants can be represented by matrices of
dimension $2N\times 2N$ ($N$ is the number of vibrational modes)
which makes the calculation of the cumulants numerically
straightforward.

In fig.\ \ref{fig_resultsC60} we show the bias-dependence of the
first three cumulants for parameters corresponding to fig.\ 3
of\cite{mccarthy_2003}. Since $\Gamma^{(0)}\ll K$, the oscillator
is in equilibrium and, therefore, the FCS of the model can also be
calculated from a simple two-level model with 4 rates given by
$P(E)$-theory. We have verified that the semi-analytical results
coincide with numerics (not shown) which we view as a non-trivial
test of our method. In non-equilibrium cases there are no simple
alternatives to the numerics. We demonstrate the full power of the
method in the second example.

\section{Model 2: The Quantum Shuttle}

\begin{figure}
  \centering
  \includegraphics[width=0.5\textwidth]{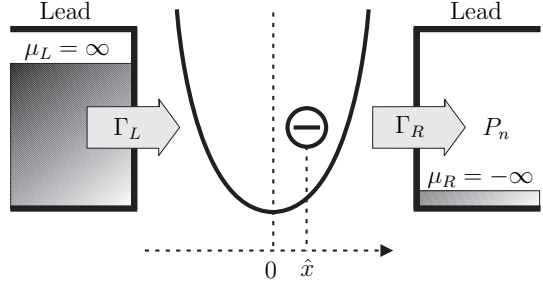}
  \caption{The quantum shuttle consists of a
  nanosized grain moving in a harmonic potential between two leads.
  A high bias between the leads drives electrons through the grain.}
  \label{fig_setup}
\end{figure}

We consider the model of a quantum shuttle used in
\cite{novotny_2003, novotny_2004, fedorets_2004}. The system
consists of an oscillating nanoscopic grain coupled to two leads
(fig.\ \ref{fig_setup}). In the strong Coulomb blockade regime the
grain effectively has just one electronic level. The oscillations
of the grain are treated fully quantum mechanically, and damping
of the oscillations is due to a surrounding heat bath. As
in\cite{novotny_2004} we consider the $n$-resolved system density
matrices $\hat{\rho}_{ii}^{(n)}(t),i=0,1$, where $n$ is the number
of electrons that have tunneled into the right lead in the time
span $t$. In the high bias limit $n$ is nonnegative, and the
$\hat{\rho}_{ii}^{(n)}(t)$ evolve according to the $n$-resolved
GME
\begin{equation}
\begin{split}
\dot{\hat{\rho}}_{00}^{(n)}(t)=&
\frac{1}{i\hbar}[\hat{H}_{\ab{osc}},\hat{\rho}_{00}^{(n)}(t)]
+\mathcal{L}_{\ab{damp}}\hat{\rho}_{00}^{(n)}(t)
-\frac{\Gamma_L}{2}\{e^{-\frac{2\hat{x}}{\lambda}},\hat{\rho}_{00}^{(n)}(t)\}
+\Gamma_Re^{\frac{\hat{x}}{\lambda}}\hat{\rho}_{11}^{(n-1)}(t)e^{\frac{\hat{x}}{\lambda}},\\
\dot{\hat{\rho}}_{11}^{(n)}(t)=&
\frac{1}{i\hbar}[\hat{H}_{\ab{osc}}-eE\hat{x},\hat{\rho}_{11}^{(n)}(t)]
+\mathcal{L}_{\ab{damp}}\hat{\rho}_{11}^{(n)}(t)
-\frac{\Gamma_R}{2}\{e^{\frac{2\hat{x}}{\lambda}},\hat{\rho}_{11}^{(n)}(t)\}
+\Gamma_Le^{-\frac{\hat{x}}{\lambda}}\hat{\rho}_{00}^{(n)}(t)e^{-\frac{\hat{x}}{\lambda}},
\end{split}
\label{eq_ngme}
\end{equation}
with $n=0,1,\ldots$ and $\hat{\rho}_{11}^{(-1)}(t)\equiv 0$. Here
the commutators describe the coherent evolution of the charged
($\rho_{11}$) or empty ($\rho_{00}$) shuttle which is modelled by
a quantum mechanical harmonic oscillator of mass $m$ and frequency
$\omega$. The electric field between the leads is denoted $E$. The
terms proportional to $\Gamma_{L/R}$ describe transfer processes
from the left/to the right lead with hopping amplitudes that
depend exponentially on the position $\frac{\hat{x}}{\lambda}$,
where $\lambda$ is the electron tunneling length. The mechanical
damping of the oscillator is described by the damping kernel (here
$T=0$) $\mathcal{L}_{\rm{damp}}\hat{\rho} =
-\frac{i\gamma}{2\hbar}[\hat{x},\{\hat{p},\hat{\rho}\}]-\frac{\gamma
m\omega}{2\hbar}[\hat{x},[\hat{x},\hat{\rho}]]$\cite{novotny_2003,novotny_2004}.
We identify the current superoperators from (\ref{eq_ngme}):
$\mathcal{I}^{+}\hat{\rho}=\Gamma_Re^{\frac{\hat{x}}{\lambda}}|0\rangle\!\langle
1|\hat{\rho}|1\rangle\!\langle
0|e^{\frac{\hat{x}}{\lambda}},\,\mathcal{I}^{-}\equiv 0$.

In\cite{novotny_2003,novotny_2004} it was found that the quantum
shuttle exhibits a crossover from tunneling to shuttling when the
damping, starting above a certain threshold value, is decreased.
This transition is clearly recorded both in the
current\cite{novotny_2003} and the zero-frequency current
noise\cite{novotny_2004}. The FCS in the tunneling and shuttling
limit is to a first approximation captured by the results for the
zero amplitude (with appropriately renormalized rates) and the
large (shuttling) amplitude of a driven
shuttle\cite{pistolesi_2004}, respectively. When approaching the
semi-classical regime, a giant enhancement of the noise was found
in the transition region. This behavior was tentatively attributed
to amplitude fluctuations in the spirit of\cite{isacsson_2004},
however, a more quantitative explanation has been missing. The
phase space representation of the stationary state of the shuttle
in the transition region indicated that shuttling and tunneling
processes coexist\cite{novotny_2004} leading  to the conjecture
that the giant noise enhancement is caused by switching between
two current channels\footnote{ Such a behavior, referred to as the
`whistle' effect, was first reported in\cite{sukhorukov_2001}.}
(tunneling and shuttling) induced by infrequent jumps between two
discrete values of the shuttle amplitude. Very recently the FCS of
such bistable systems has been studied\cite{jordan_2004}, and it
was found that the first three cumulants are (assuming that the
individual channels are noiseless)
\begin{subequations}
\begin{align}
\llangle I\rrangle
&=\frac{I_{\rm{S}}\Gamma_{\rm{S}\leftarrow\rm{T}}
+I_{\rm{T}}\Gamma_{\rm{T}\leftarrow\rm{S}}}
{\Gamma_{\rm{T}\leftarrow\rm{S}}+\Gamma_{\rm{S}\leftarrow\rm{T}}}\
, \label{eq_firstan}\\ \llangle I^2\rrangle
&=2(I_{\rm{S}}-I_{\rm{T}})^2\frac{\Gamma_{\rm{S}\leftarrow\rm{T}}\Gamma_{\rm{T}\leftarrow\rm{S}}}{(\Gamma_{\rm{S}\leftarrow\rm{T}}
+\Gamma_{\rm{T}\leftarrow\rm{S}})^3}\ ,\label{eq_secondan}
\\ \llangle I^3\rrangle&=
6(I_{\rm{S}}-I_{\rm{T}})^3\frac{\Gamma_{\rm{S}\leftarrow\rm{T}}\Gamma_{\rm{T}\leftarrow\rm{S}}
(\Gamma_{\rm{T}\leftarrow\rm{S}}-\Gamma_{\rm{S}\leftarrow\rm{T}})}
{(\Gamma_{\rm{S}\leftarrow\rm{T}}+\Gamma_{\rm{T}\leftarrow\rm{S}})^5}\
.\label{eq_thirdan}
\end{align}
\end{subequations}
Here $I_{\rm{S}/\rm{T}}$ denote the current associated with the
shuttling/tunneling channel\footnote{ $I_{\rm{S}}=\omega/2\pi$,
$I_{\rm{T}}=\frac{\tilde{\Gamma}_L\tilde{\Gamma}_R}{\tilde{\Gamma}_L+\tilde{\Gamma}_R}$,
with $\tilde{\Gamma}_R=\Gamma_R
e^{\hbar/m\omega\lambda^2}e^{2eE/m\omega^2\lambda}$,
$\tilde{\Gamma}_L=\Gamma_L e^{\hbar/m\omega\lambda^2}$
\cite{novotny_2004,donarini_2004}.}, while
$\Gamma_{\rm{T}\leftarrow\rm{S}}$ is the transition rate from the
shuttling to the tunneling channel and
$\Gamma_{\rm{S}\leftarrow\rm{T}}$ is the rate of the reverse
transition.

The rates can be evaluated analytically in the limit
$\lambda\to\infty$ and $E\to 0$ \cite{donarini_2004}. However, for
other parameters an alternative approach is needed. First, one
evaluates the first three cumulants numerically following the
theory presented above. Then, one calculates the rates from the
first two cumulants (eqs.\ (\ref{eq_firstan},\ref{eq_secondan}))
and compares the numerically calculated third cumulant from eq.\
(\ref{eq_thirdcumulant}) with the one obtained from eq.\
(\ref{eq_thirdan}). The numerical calculation of the cumulants is
only possible by a truncation of the oscillator Hilbert space. By
retaining the $N$ lowest oscillator states the (non-sparse) matrix
representations of the superoperators entering eqs.\
(\ref{eq_firstcumulant}-\ref{eq_thirdcumulant}) are of dimension
$2N^2\times 2N^2$, which for the required values of $N\sim 100$
leaves us with non-trivial numerical matrix problems. However,
using the iterative methods described in\cite{flindt_2004} the
cumulants can be evaluated numerically.

\begin{figure}
  \centering
  \includegraphics[width=0.95\textwidth]{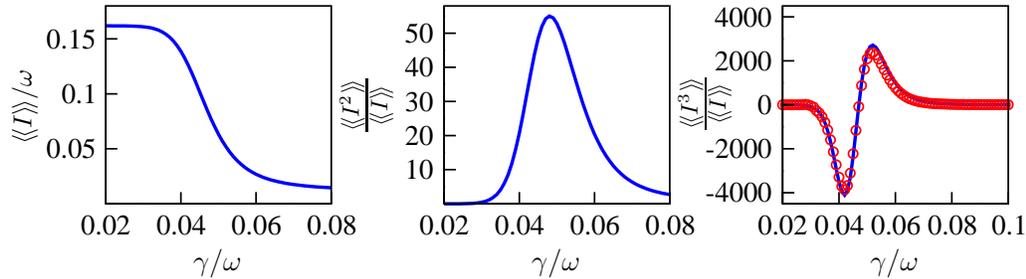}
  \caption{Results for the quantum shuttle. First three cumulants
   as function of the damping $\gamma$. The parameters are
   $\Gamma_L=\Gamma_R=0.01\omega$, $\lambda
   =1.5x_0$, $d\equiv eE/m\omega^2=0.5x_0$, $x_0=\sqrt{\hbar/m\omega}$.
   The full lines indicate numerical results, while the circles indicate
   the analytic expression for the third cumulant assuming that
   the shuttle in the transition region effectively behaves as a bistable system.}
  \label{fig_resultsshuttle}
\end{figure}

If fig.\ \ref{fig_resultsshuttle} we show the $\gamma$-dependence
of the first three cumulants for $\lambda= 1.5x_0$, where
$x_0=\sqrt{\hbar/m\omega}$. The first cumulant, the current, shows
the transition from the tunneling to shuttling current with
decreasing damping. The transition is also evident from the second
cumulant, the zero-frequency current noise, which exhibits a giant
enhancement in the transition region, before dropping to very low
values in the shuttling region. Together with the numerical
results for the third cumulant we show the analytic expression
(eq.\ (\ref{eq_thirdan})) with rates extracted from the first two
cumulants. As can be seen the two  data sets coincide, which we
take as evidence that the quantum shuttle in the transition region
indeed behaves as a bistable system for which the FCS is
known\cite{jordan_2004}. When approaching the deep quantum regime,
$\lambda\sim x_0$ (not shown), the transition from tunneling to
shuttling is smeared out and the distinct current channels cease
to exist. In this limit the bistable system model is not valid.

\section{Conclusion}

We have presented a method for computation of the FCS for typical
nanoelectromechanical systems and applied it to two specific
models. For the \chem{C_{60}}-setup with equilibrated oscillator
we have calculated the first three cumulants and explained the
results in terms of a simple two-level model. For the quantum
shuttle we have used the first three cumulants as evidence that
the shuttle in the transition region behaves as a bistable system.
This example clearly illustrates the usefulness of the FCS in
probing a microscopic system. Here we have only shown explicit
expressions for the first three cumulants, our method, however,
can be extended to the calculation of cumulants of any order, and
we believe that the method has a broad range of applicability.

\acknowledgments The authors would like to thank A. Armour and A.
Donarini for careful reading of the manuscript and fruitful
discussions.

\end{document}